\documentclass[preprint, 12pt]{aastex62}
\usepackage[T1]{fontenc}

\usepackage{bm}        
\usepackage{amssymb, amsmath}   
\usepackage{cancel}
\usepackage{color}
\usepackage{comment}

\newcommand{\vect}[1]{\bm{#1}}

\renewcommand{\thefootnote}{\arabic{footnote}}

\def\solphys{Sol.~Phys.}      
\def\apj{Astrophysical Journal}                 
\def\apjl{Astrophysical Journal}                 
\def\mnras{MNRAS}                 
\def\ssr{Space Science Reviews}                 
\def\physrep{Physics Reports}                 
\def\aap{Astronomy and Astrophysics}                 

\DeclareMathAlphabet{\mathsfit}{\encodingdefault}{\sfdefault}{m}{sl}
\SetMathAlphabet{\mathsfit}{bold}{\encodingdefault}{\sfdefault}{bx}{sl}
\DeclareMathOperator{\sech}{sech}

\begin{document}



\title{Recent Progress on Particle Acceleration and Reconnection Physics during Magnetic Reconnection in the Magnetically-dominated Relativistic Regime}

\author{Fan Guo}
\affiliation{Los Alamos National Laboratory, NM 87545 USA}
\author{Yi-Hsin Liu}
\affiliation{Dartmouth College, Hanover, NH 03750 USA}
\author{Xiaocan Li}
\affiliation{Dartmouth College, Hanover, NH 03750 USA}
\author{Hui Li}
\affiliation{Los Alamos National Laboratory, NM 87545 USA}
\author{William Daughton}
\affiliation{Los Alamos National Laboratory, NM 87545 USA}
\author{Patrick Kilian}
\affiliation{Los Alamos National Laboratory, NM 87545 USA}
\date{\today}

\begin{abstract}
Magnetic reconnection in strongly magnetized astrophysical plasma environments is believed to be the primary process for fast energy release and particle energization. Currently there is strong interest in relativistic magnetic reconnection, in that it may provide a new explanation for high-energy particle acceleration and radiation in strongly magnetized astrophysical systems.
We review recent advances in particle acceleration and reconnection physics in the magnetically-dominated regime. More discussion is focused on the physics of particle acceleration, power-law formation as well as the reconnection rate problem. In addition, we provide an outlook for studying reconnection acceleration mechanisms and kinetic physics in the next step.

\end{abstract}



\section{Introduction} 

Magnetic reconnection is a fundamental plasma process that breaks and rejoins magnetic field lines across a magnetic shear. In a strongly magnetized plasma, magnetic reconnection liberates a large amount of magnetic energy and drives bulk flows, plasma heating, and particle acceleration. Magnetic reconnection is a long-standing research topic. Previously, most reconnection studies have focused on laboratory \citep{Ji1998,Egedal2011}, space \citep{Phan2000,Birn2001,Hoshino2001,Fu2011,Birn2012,Wang2016coalescence}, or solar environments \citep{Kopp1976,Gordovskyy2010,Tian2014,Chen2018,Chen2020}. Recently, there is a strong surge of interest in relativistic magnetic reconnection, as it may be a prodigious source of nonthermal particles and emissions in a rich set of high-energy astrophysical activities. For example, in plasma environments associated with compact objects such as pulsars, magnetars, black holes and their binary and coalescence systems, the magnetic field can be extremely large and becomes important, even dominant, in the energetics and plasma dynamics of the system. Magnetic field can play a significant role in pulsar wind nebulae (PWN) \citep{Coroniti1990,Kirk2003,Arons2012,Hoshino2012} and jets in gamma-ray bursts (GRBs) \citep{Drenkhahn2002,Zhang2011,McKinney2012} and from black holes \citep{Pino2005,Giannios2009,Zhang2015,Zhang2018,Bottcher2019}. The launched relativistic flows are likely Poynting-flux dominated, meaning the magnetization parameters $\sigma$ (the ratio of the magnetic energy density to the plasma energy density $\sigma =B^2/(8\pi w)$, where $w$ is the enthalpy) can be much greater than unity and the Alfv\'en speed approaches the speed of light $V_A \sim c$. Those systems have a rich variety of flares and bursty phenomena, featured by an explosive unleash of energy and the associated increase of energetic particles and emissions. Remarkable examples are Crab flares
\citep{Tavani2011,Abdo2011}, gamma-ray bursts \citep{Zhang2011,McKinney2012,Kumar2015}, magnetar flares \citep{Thompson2002,Lyutikov2003,Palmer2005}, and TeV Blazar emission \citep{Hayashida2015,Ackermann2016,Yan2015,Yan2016}. In the magnetically dominated scenarios, magnetic reconnection is thought to be the driver for energy release and particle acceleration. To explain the high-energy emissions, it is generally expected that the accelerated particles (electrons and ions) should develop a nonthermal power-law energy distribution extending to high energy \citep{Matthews2020}.

There has been a strong interest in relativistic reconnection over the past decades in plasma astrophysics \citep{Blackman1994,Lyutikov2003,Lyubarsky2005,Comisso2014,Sironi2014, Guo2014,Guo2015,Takamoto2013,Liu2015,Liu2017,Liu2020}, but its rich physics and its associated particle acceleration in the relativistic regime remain less studied compared to the non-relativistic counterparts. While several analytical models have been proposed for reconnection rate and particle acceleration, recent studies of particle-in-cell simulations have substantially explored this regime. 
Since the magnetic field is the source of free energy for particle energization processes during magnetic reconnection, it is useful to define several parameters for  comparing the magnetic energy with other characteristic plasma energies.
The first is the magnetization parameter (without finite temperature effects) $\sigma_0 = B_0^2 / (8 \pi \rho \mathrm{c}^2)$ where $B_0$ is the magnetic field strength, $\rho$ is mass density and $\mathrm{c}$ is the speed of light. This ratio between energy density in the magnetic field to the energy density associated with the rest mass of the plasma can also be seen as the amount of magnetic energy potentially available per particle.
The second parameter is the plasma $\beta$ defined as $\beta = 8 \pi n k_B T /B_0^2$. This well known plasma parameter compares the thermal pressure of the gas with the magnetic pressure.
Alternatively we can define $\sigma_\mathrm{th} = B_0^2/(12 \pi n k_B T) = 2/(3\,\beta)$, the ratio between magnetic field energy and thermal energy density that measures the maximum possible energization per particle compare with the thermal energy \citep{Kilian2020}. 

There are two outstanding problems that drive theoretical investigations of energy release and particle acceleration in reconnection, namely:  What determines the time scale of magnetic energy release? And how particles are accelerated to high energy? These are not only fundamental problems of magnetic reconnection, but also critical for explaining energy release and the outcome of particle energization in high-energy astrophysics phenomena.
The past several years have
seen a number of significant advances regarding the reconnection physics and particle acceleration
mechanisms in the relativistic regime. It is discovered that in the relativistic reconnection regime strong
nonthermal particle acceleration occurs and the resulting particle energy spectra resemble power-law distributions $f \propto \gamma^{-p}$ \citep{Sironi2014,Guo2014,Guo2015,Werner_2015, Guo2019}. Meanwhile, the normalized inflow speed in the relativistic collisionless regime is $R \sim 0.1-0.2$ times the Alfv\'en speed, similar to studies in the nonrelativistic regime, indicating fast and efficient energy conversion \citep{Liu2015,Werner_2017a,Liu2017,Liu2020}. 
These suggest that relativistic magnetic reconnection is a promising scenario for explaining energy release and nonthermal particle acceleration in high-energy astrophysics.
Establishing the importance of relativistic magnetic reconnection in astrophysical high-energy processes requires further understanding of these theoretical problems: 1.  What physics determines the fast energy release? What is the reconnection rate and what set of physics for determining the rate? 2. How are particles accelerated into a nonthermal power-law distribution? What are the primary acceleration mechanism and the mechanism for developing a power-law distribution?
Over the past a few years, significant progress has been made on plasma dynamics, particle acceleration, and reconnection physics in the relativistic reconnection regime. Meanwhile, other interesting questions such as 3D effects, roles of turbulence, etc. are emerging. It is a rapidly evolving field with many exciting results. This review provides an overview of some of the recent results and points out several important issues to study in the next step.

\section{Particle Acceleration in Relativistic Magnetic Reconnection}

Despite the strong anticipation and observation evidence on the role of magnetic reconnection in accelerating energetic
particles, a complete theory of particle acceleration in the reconnection region is still a work-in-progress. Much of the recent progress has been made by particle-in-cell (PIC) kinetic simulations, as it offers a self-consistent and complete description for kinetic plasma physics. Compared to fluid descriptions, fully kinetic simulations can model collisionless reconnection electric field in genearlized Ohm's law without assumptions, and thus offer a robust kinetic description of magnetic reconnection. It also self-consistently includes nonthermal particle acceleration out of the thermal pool, development of power-law distributions, and the feedback of energetic particles to the system. However, PIC simulations have to resolve relevant plasma kinetic scales, which makes it difficult to extrapolate the results to scales relevant to astrophysical observations. Nevertheless, recent PIC simulations have obtained numerous important results toward a comprehensive understanding of particle acceleration in relativistic reconnection. We will briefly discuss the development of large-scale reconnection acceleration model in Section~\ref{subsubsec:large_scale}.

Previous PIC simulations in the nonrelativistic regime have considerable difficulties in generating clear power-law energy distributions. This led to some debate in the community about the mechanism of power-law formation \citep{Drake2010,Drake2013,Spitkovsky2019}. One popular speculation is that those simulations are mostly performed in a periodic simulation domain and an ``escape'' mechanism is required to generate a power-law distribution. It is worth noting that most of the earlier nonrelativistic simulations are in the parameter range with plasma $\beta \sim 1$ ($\sigma_{th} \sim 1$), allowing only a limited amount of magnetic energy converted into plasma energy, in comparison to the initial plasma thermal energy. 
The recent activity of PIC simulations in the relativistic regime offers a crucial test to the understanding of the formation of power-law distribution in the limit of strong magnetic energy conversion and particle acceleration ($\sigma_0 \gg 1$ and $\sigma_{th} \gg 1$). As we will discuss in Section~\ref{subsec:basic_results}, this greatly changes the resulting particle distribution and our understanding of power-law formation. In the relativistic regime, magnetic reconnection develops clear power-law distributions even in a periodic simulation domain \citep{Sironi2014,Guo2014,Guo2015,Werner_2015,Guo2016}. 

While the power-law distributions from relativistic magnetic reconnection appear to be a robust result, there is no consensus established on the primary acceleration mechanism \citep{Sironi2014,Guo2014,Guo2015,Guo2019}. In Section~\ref{subsec:acceleration_mechanism} we will discuss this issue. 
In Section~\ref{subsec:powerlaw_creation} we discuss in detail the mechanism for the formation of power-law distribution. We show analytically that the formation of power-law distribution does not require an ``escape'' term (but its presence can modify the spectral index) and that an injection process is important for power-law formation. This not only clarifies the power-law formation issues in the relativistic reconnection studies, but also provides a ``game changer'' for studying power-law formation in the nonrelativistic regime, namely power-law distributions are much easier to develop in low-$\beta$ regime ($\beta \ll 1$ or $\sigma_{th} \gg 1$)). This has driven new studies in the nonrelativistic regime with low plasma $\beta$ (high $\sigma_{th}$) condition \citep{Li2015,Li2017,Li2018Role,Li2018Large,Li2019Particle,Li2019Formation}.

In Section~\ref{subsec:outlook} we provide an outlook for a number of issues important to resolve and make progress in the future. This includes the particle injection process (\ref{subsubsec:injection}), guide field dependence (\ref{subsubsec:guide_field}), 3D effects (\ref{subsubsec:3d_physics}), and the pathway toward a large-scale theory (\ref{subsubsec:large_scale}).

\subsection{Basic results}
\label{subsec:basic_results}

\begin{figure}
\centering
\includegraphics[width=0.6\textwidth]{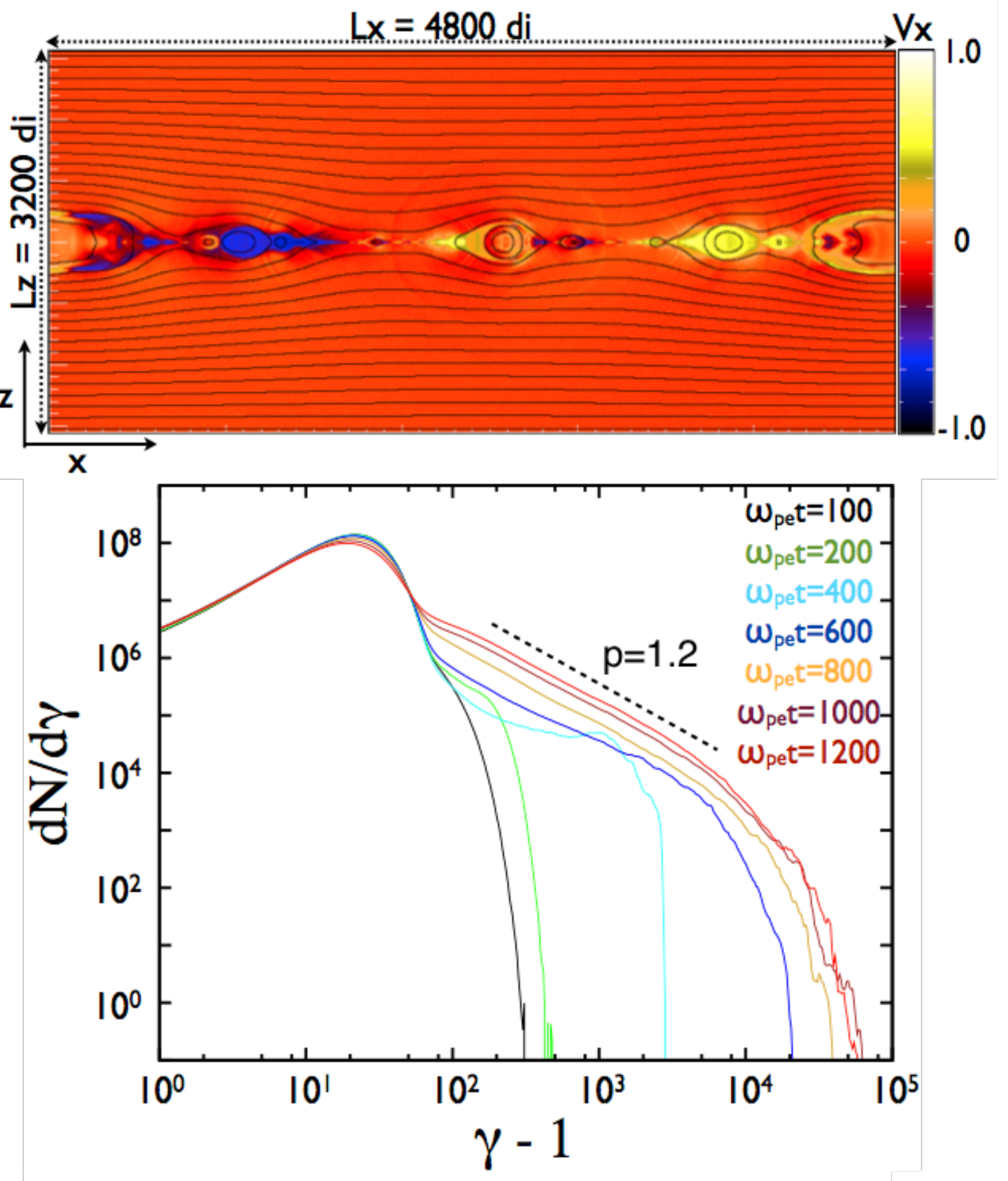}
\caption{Dynamics and resulting energy spectrum in relativistic pair plasma reconnection from a sample PIC simulation starting from a force-free current sheet with $\sigma_0 = 3200$. Top: A snapshot at $\omega_{pe}t = 1200$ showing the outflow speed normalized by the speed of light. Bottom: energy spectra at different time steps generated from the simulation.  \label{fig:pic_run} }
\end{figure}

\begin{figure}
\centering
\includegraphics[width=0.85\textwidth]{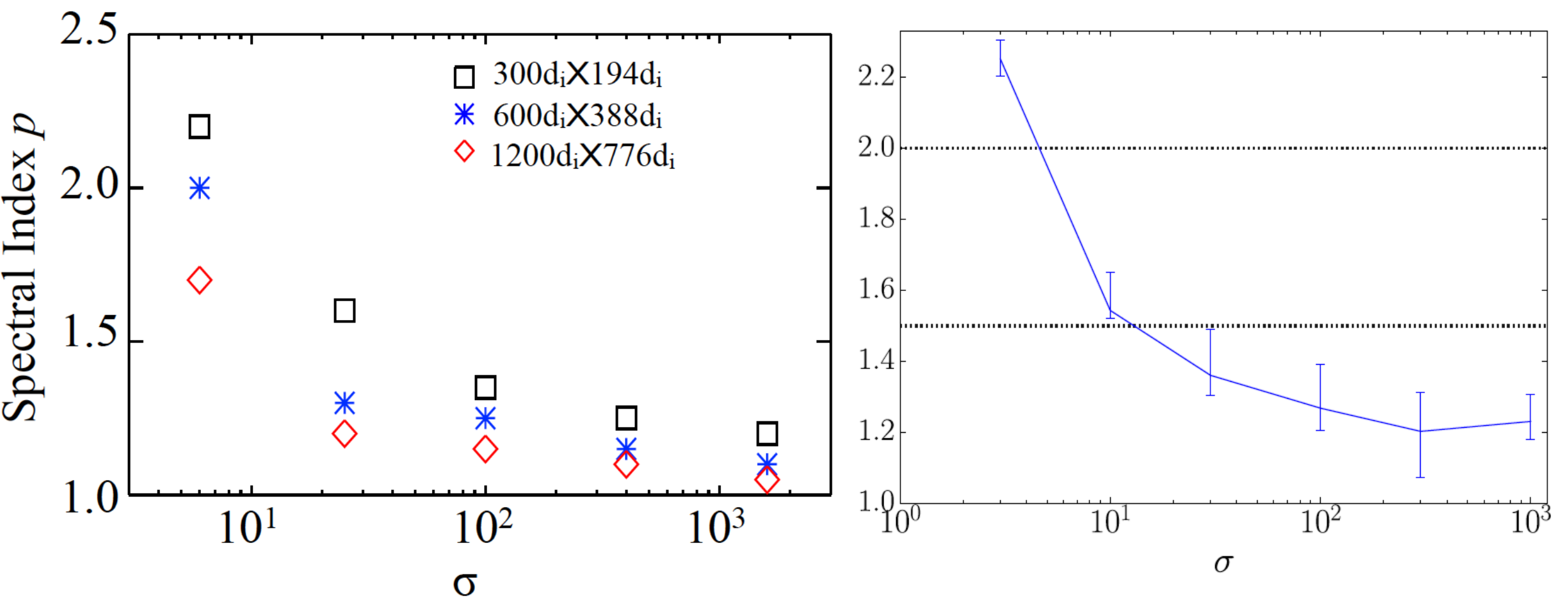}
\caption{Spectral index as a function of $\sigma$ for different box sizes from two studies. Adapted from \citet{Guo2014} and \citet{Werner_2015}. \label{fig:power_sigma} }
\end{figure}

Perhaps the most interesting result in particle energization during relativistic magnetic reconnection is the development of power-law distributions. Several different numerical studies for pair plasmas have shown that, in the relativistic regime ($\sigma_0 > 1$), magnetic reconnection is efficient at accelerating particles into relativistic energies. Much of the magnetic energy is converted into the kinetic energy of nonthermal relativistic particles and the eventual energy spectra resemble a power-law $f(\gamma) \propto \gamma^{-p}$. The spectra are harder for higher $\sigma_0$ and the spectral index approaches $p \sim 1$ for sufficiently large $\sigma_0$ and system size, and the break energy is at least several times $\sigma_0$ \citep{Sironi2014,Guo2014,Guo2015,Werner_2015}.
Fig.~\ref{fig:pic_run} shows a sample 2D PIC simulation in the $x$--$z$ plane using the VPIC code \citep{Bowers2008}. The simulation starts from a magnetically-dominated force-free current sheet $\vect{B}=B_0\tanh(z/\lambda)\hat{x}+B_0\sqrt{\sech^2(z/\lambda)+B_g^2/B_0^2}\hat{y}$, where $B_g$ is the strength of the guide field (setting to zero in this case). The plasma consists of electron-positron pairs with mass ratio $m_i = m_e$ and $\sigma_0 = 3200$. The domain size is $L_x \times L_z = 4800 d_i \times 3200 d_i$, where $d_i$ is the electron skin depth. The boundary conditions for 2D simulations are periodic for both fields and particles in the x-direction, while in the z-direction the boundaries are conducting for the field and reflecting for the particles. Readers are referred to our earlier publications for more simulation details \citep{Guo2014,Guo2016,Guo2016PoP,Kilian2020}. During magnetic reconnection, the current layer quickly breaks into several fast-moving secondary plasmoids. The plasmoids coalesce and eventually merge into a single island due to the periodicity. This basic 2-D picture has been confirmed in many studies \citep[e.g.,][]{Daughton2007,Sironi2014,Guo2015,Liu2015,Kilian2020}. A significant power-law distribution with $p\sim 1.2$ develops as reconnection proceeds. Fig.~\ref{fig:power_sigma} shows two studies on the spectral index $p$ as a function of $\sigma_0$ for different box sizes obtained from different simulations \citep{Guo2014,Werner_2015}. These clearly show that the spectral index $p$ is close to $1$ in the limit of large $\sigma_0$. 
There have been several 3D simulations performed to examine the role of 3D instabilities on particle acceleration. The pioneering work by \citet{Zenitani2008} concluded that the rapid growth of the relativistic drift kink instability deforms a current sheet without a guide field and thus prohibits particle acceleration. 
Recent 3D simulations of anti-parallel reconnection show that nonthermal acceleration can still operate in the nonlinear stage in 3D reconnection and nonthermal power-law distribution still develops \citep{Sironi2014,Guo2014,Guo2015,Werner_2017b}.
We discuss the 3D dynamics in the following sections. 

There have been additional studies on magnetic reconnection in proton-electron plasmas \citep{Guo2016,Werner_2017a,Ball_2018}. For a proton-electron plasma with a total magnetization of $\sigma_0$, the magnetization for proton is $\sigma_p \sim \sigma_0$ and $\sigma_e \sim \sigma_0 m_i/m_e$, respectively. For the case of $\sigma_p \gg 1$, the simulation results are similar to results from the pair plasma simulation with $1 < p < 2$ \citep{Guo2016}. Alternatively, \citet{Werner_2017a} have suggested that the limit to $\sigma_0 \gg 1$ is $p \sim 2$. However, note that this limit is for a fixed $\beta = 0.02$, not toward the most extreme regime where $\sigma_p \gg 1 $ and $\beta \ll 1$.

Recently, \citet{Petropoulou2018} studied the long-term evolution of energy spectrum in large two-dimensional kinetic simulations of relativistic reconnection and found the break energy sustainably increases and spectrum continuously softens. We note that effects like particle loss from more realistic boundary conditions as well as 3D effects would be important to consider. In 2D systems, particles can be artificially confined in large magnetic islands, limiting the acceleration of high-energy particles \citep{Li2019Formation,Dahlin2017}. We discuss this issue more in the following sections. 

\subsection{Particle Acceleration Mechanism}
\label{subsec:acceleration_mechanism}

Driven by the discovery of power-law energy spectra in relativistic magnetic reconnection, there have been active discussions on how particles are accelerated to high energy. While a growing set of research has suggested that relativistic magnetic reconnection can be an efficient source of particle acceleration in high-energy astrophysical systems, the dominant acceleration mechanism remains controversial. 
Past research on particle acceleration during magnetic reconnection has mainly identified two mechanisms: a Fermi-type acceleration mechanism where particles are accelerated by bouncing back and forth in the reconnection generated bulk flows \citep{Kliem1994,Pino2005,Drake2006,Fu_2006,Drury_2012,Guo2014,Guo2015,Guo2019,Dahlin2014,Zank2014,LeRoux2015,Li2018Role,Li2018Large,Li2019Formation} and direct acceleration at diffusion regions surrounding reconnection X-points \citep{Ambrosiano1988,Litvinenko1996,Zenitani_2001,Pritchett_2006,Oka2010,Wang_2016,Sironi2014}. The Fermi-type acceleration is mainly through the electric field induced by bulk plasma motion $\vec{E}_m = - \vec{u} \times \vec{B}/c$ perpendicular to the local magnetic field, whereas the direct acceleration is driven by the parallel electric field if a non-zero magnetic field exists, or through Speiser orbits in the nonideal electric field when the magnetic field is weak \citep{Speiser1965}. It is therefore useful to distinguish the relative contribution of the two during the particle acceleration process, either according to generalized Ohm's law \citep{Guo2019}, or simply by decomposing the electric field into the perpendicular part $E_\perp$ and parallel component $E_\parallel$ and evaluating the work done by each of them \citep{Guo2015,Ball_2019,Kilian2020}. One can further decompose the particle motion into various drift motions based on the guiding-center approximation, which captures most of the particle energization processes in the strongly magnetized regime. While analyzing single particle trajectories is useful in terms of identifying basic acceleration patterns, in the past this has generated significant controversy and confusion about the relative importance of the two mechanisms, as the presented trajectories are subject to ``cherry-picking''. It is therefore very important to statistically study the acceleration mechanisms and consider all possibilities without bias. Recently, there have been more PIC studies showing acceleration mechanisms statistically in reconnection and other scenarios, and most of them point toward the importance of Fermi mechanisms in accelerating particles to high energy \citep{Guo2014,Guo2015,Guo2019,Comisso2018,Alves2018,Li2018Role,Li2019Particle,Li2019Formation,Kilian2020}.

\begin{figure}
\centering
\includegraphics[width=0.9\textwidth]{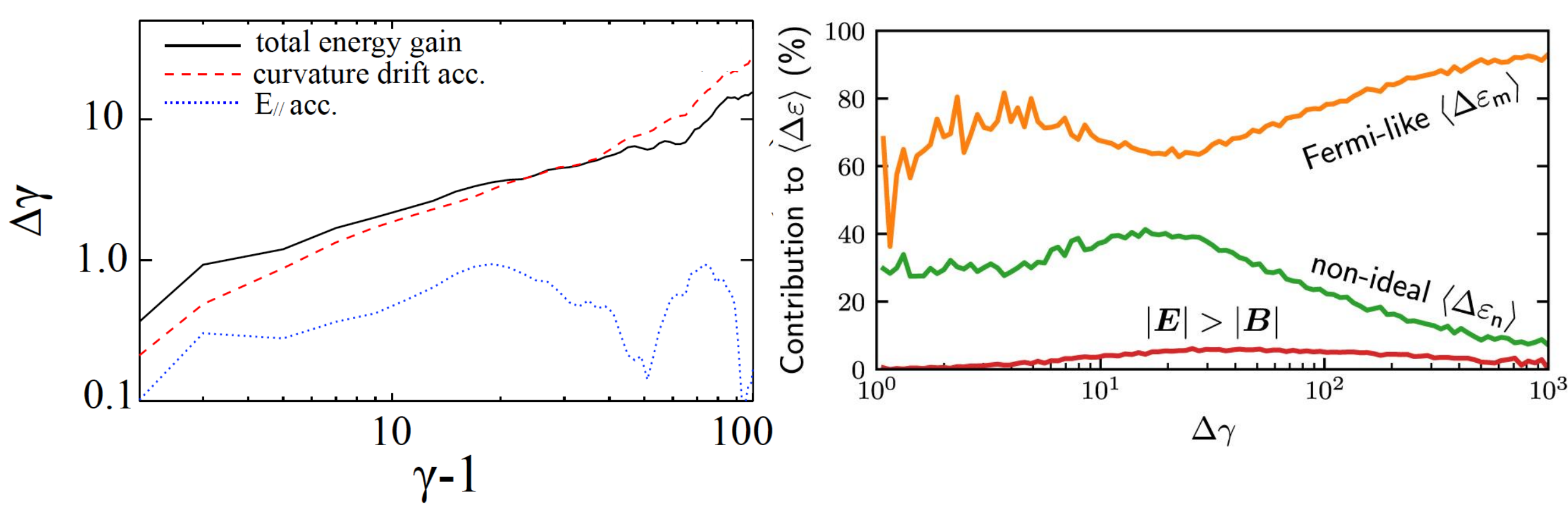}
\caption{Statistics of particle acceleration for evaluating the contribution of different mechanisms. Left: Averaged energy gain and contributions
from parallel electric field acceleration and curvature drift acceleration
over an interval of $25\omega_{pe}^{-1}$ as a function of particle energy at a simulation time, adapted from \citet{Guo2014}; Right panel shows statistics of different acceleration mechanism for $\sim 1$ million particles traced over the history of the simulation as a function of energy gain till the end of simulation, adapted from \citet{Guo2019}. The orange line shows the fraction of averaged energy gain from motional electric field. The green line shows the contribution of the non-ideal electric field and the red line shows the contribution of electric field in regions with $|E| > |B|$, as suggested by \citet{Sironi2014}. The acceleration to high-energy is dominated by the Fermi-type acceleration process. \label{fig:mechanism} }
\end{figure}

In 2D simulations, it was found, through several different analyses, that a Fermi-like acceleration driven by plasmoid motion dominates the acceleration process \citep{Guo2014,Guo2015,Guo2019} (for a nonrelativistic description see \citep{Dahlin2014,Zhou2015,Li2017,Li2018Role,Li2019Formation}) in the weak guide field regime. This main acceleration mechanism is supported by curvature drift motion along the direction of electric field induced by plasma flows. In PIC simulations of magnetic reconnection in low plasma $\beta$ and weak guide field, strong compression leads to fast reconnection energy conversion and particle acceleration \citep{Li2018Role}. 
Fig.~\ref{fig:mechanism} shows a representative example of statistical results on the acceleration mechanisms, distinguishing the roles of Fermi acceleration and parallel electric field acceleration, adapted from \citep{Guo2014,Guo2019}. The left panel shows that the energy gain by the Fermi acceleration supported by curvature drift acceleration along the perpendicular electric field, parallel electric field, and total energy gain in a time interval around the time that reconnection rate peaks in a 2D PIC simulation of relativistic reconnection. 
It shows that the main energy gain is through Fermi acceleration, and the energy gain in Fermi acceleration is proportional to the particle energy $\delta \gamma \propto \gamma - 1$, i.e., the acceleration rate $\alpha = \dot{\varepsilon}/\varepsilon$ is nearly a constant, recovering the classical Fermi acceleration. This analysis was further done in a more sophisticated way in 3D simulations by \citet{Li2019Formation}. The constant acceleration rate as a function of particle energy is an important ingredient for acceleration into a power-law distribution. The right panel shows the fraction of averaged energy gain as a function of energy for particles at the end of the simulation. At intermediate energies, the energy gain from Fermi acceleration is comparable to the non-ideal electric field. For particles accelerated to high energy, the Fermi-like process dominates the acceleration process. 
On the other hand, the acceleration through parallel electric field is subdominant \citep{Guo2014,Guo2015,Guo2019}. In addition, the acceleration through X-points with $|E|>|B|$ (suggested by \citet{Sironi2014}) is very small.

\subsection{Development of Power-law Energy Distribution}
\label{subsec:powerlaw_creation}

To explain emissions from high-energy astrophysical processes, it is important to see whether reconnection is able to generate a power-law particle energy distribution. 
Even though the main particle acceleration mechanism is correctly identified in previous nonrelativistic studies \citep{Dahlin2014,Li2015,Li2017}, it is still not clear if any power-law energy spectrum can be identified from those simulations. This suggests that formation of power-law energy distribution requires a more restricted condition that is not achieved in those simulations. As discussed above, recent studies have shown that power-law distribution may be a common form of particle energy spectrum in relativistic magnetic reconnection. These new results have offered an opportunity to study the condition of power-law particle distributions generated during magnetic reconnection.  

\begin{figure}
\centering
\includegraphics[width=0.9\textwidth]{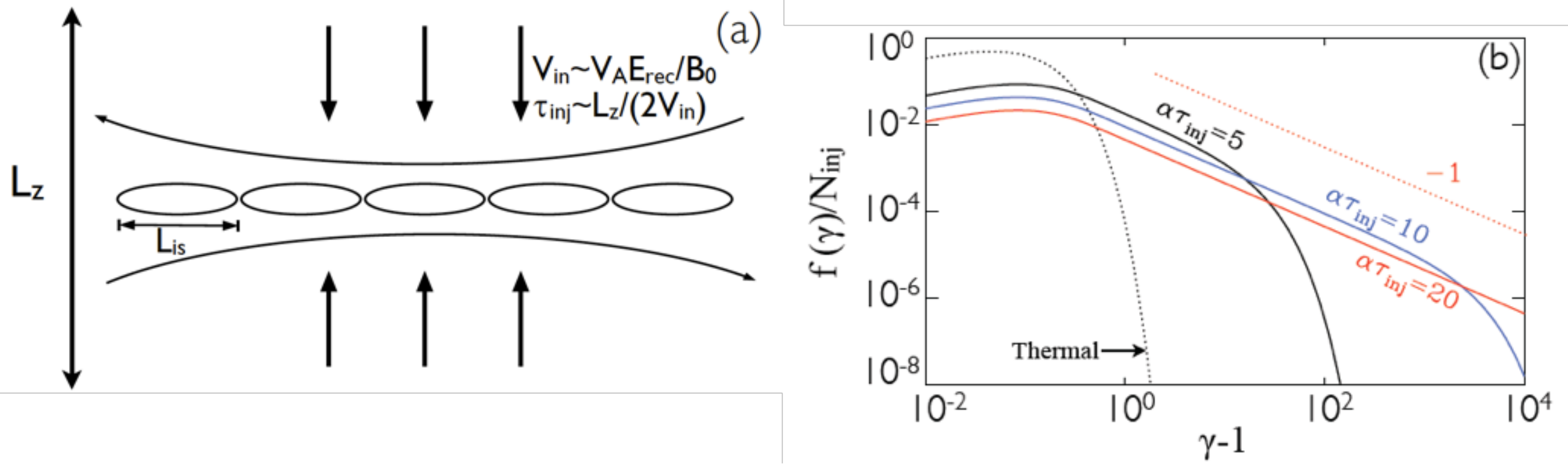}
\caption{(a) Geometry of a simple acceleration model for describing the formation of power-law distribution under the influence of injection and Fermi acceleration in the reconnection layer. (b) Analytical distribution function of accelerated particles for different values of $\alpha \tau_{inj}$ from Equation (6). Both adapted from \citet{Guo2014} \label{fig:example} }
\end{figure}

To explain the power laws of particle energy distribution observed in PIC simulations, \citet{Sironi2014} have proposed that the power-law form is established as the particles accelerated at the X-points (diffusion regions with weak magnetic field $|E| > |B|$ when the guide field is zero) through direct acceleration, following \citet{Zenitani_2001}. They argue that this process is essential for the formation of power-law distributions and it determines the spectral index of the energy spectra integrated over the whole simulation domain. In contrast, \citet{Guo2014,Guo2015} have proposed that the power-law distributions are produced by a Fermi-like process and continuous injection from the reconnection inflow. This model is illustrated in Fig.~\ref{fig:example}. During reconnection particles enter the reconnection region as an inflow and get accelerated. The main acceleration in 2D reconnection systems is by Fermi acceleration supported by plasmoids. Based on this idea, they developed a simple theoretical model that is consistent with the hard spectra $f \propto \varepsilon^{-p}$ observed in the simulations, i.e., spectral index approaching $p = 1$. As we discussed in Section~\ref{subsec:acceleration_mechanism}, various studies have shown that Fermi acceleration dominates the acceleration of particles to high energy.  \citet{Guo2019} showed that the energy spectrum for particles in the diffusion regions with strong non-ideal electric field is not the same as the spectrum integrated over the whole simulation domain. While the non-ideal electric field can act as an additional particle injection for further Fermi acceleration, it is not necessary for the formation of power-law distributions. Particles that went through the diffusion regions have the same spectral slope as the particles that never encounter those regions, because of the Fermi acceleration.

There has been active discussions on the requirement for forming a power-law distribution \citep{Drake2010,Drake2013,Guo2014,Guo2019,Sironi2014,Spitkovsky2019,Li2019Formation}. It was usually argued that an escape mechanism is necessary for forming a power-law distribution. This statement is not correct, or at least inaccurate, therefore it is important to establish a common language on the basics of power-law formation.  Below we show this by a simple derivation. A more general equation and solution useful for studying the formation of power-law energy spectra can be found in \citet{Drury1999}.

A standard equation for studying particle acceleration is a Fokker-Planck equation:
\begin{eqnarray} 
\frac{\partial f}{\partial t} + \frac{\partial}{\partial \varepsilon} (\dot{\varepsilon} f) - \frac{\partial}{\partial \varepsilon} (D_{\varepsilon\varepsilon} \frac{\partial f}{\partial \varepsilon}) = \frac{f_{inj}}{\tau_{inj}} - \frac{f}{\tau_{esc}}
\end{eqnarray}

\noindent where $\dot{\varepsilon}$ can include both first-order effect and second-order effect $\dot{\varepsilon} = A - \partial D_{\varepsilon \varepsilon} / \partial \varepsilon $ \citep{Petrosian2016}. We have normalized energy as their initial thermal kinetic energy $\varepsilon = m_e c^2 (\gamma - 1)/(k_BT)$. 
To obtain a useful analytical solution for discussing the formation of power-law distributions, we 
consider the simplified energy continuity equation compare to Equation (1) \citep{Guo2014,Guo2015,Guo2019,Drury_2012}:
\begin{eqnarray} 
\frac{\partial f}{\partial t} + \frac{\partial}{\partial \varepsilon} (\dot{\varepsilon} f) = \frac{f_{inj}}{\tau_{inj}} - \frac{f}{\tau_{esc}}
\end{eqnarray}

\noindent We assume $\dot{\varepsilon} = \alpha \varepsilon$ for a Fermi-type acceleration, as supported by statistical analysis on particle acceleration. 
The energy continuity equation (2) can then be written as 
\begin{eqnarray} 
\frac{d f}{d t} +  \left(\alpha + \frac{1}{\tau_{esc}} \right)f = \frac{f_{inj}}{\tau_{inj}} 
\end{eqnarray}

 One can use Eq. (3) to study effects of acceleration, escape, and injection. In the simplest case, we ignore effects of escape ($\tau_{esc} \rightarrow \infty $) and injection ($\tau_{inj}  \rightarrow \infty$) and we assume the initial particle distribution is a thermal nonrelativistic Maxwellian distribution $f_0 = \frac{2N_0}{\sqrt{\pi}}\sqrt{\varepsilon}\exp{(-\varepsilon)}$. The energy spectrum after time $t$ then is: 
\begin{eqnarray} 
f (\varepsilon, t) = \frac{2N_0}{\sqrt{\pi}} \sqrt{\varepsilon} e^{-3\alpha t/2} \exp(-\varepsilon e^{- \alpha t}),
\end{eqnarray}

\noindent which remains a Maxwellian distribution with a temperature $e^{\alpha t} T$. However, it is trivial to show that even if we include an escape term with finite $\tau_{esc}$, the spectrum is still a Maxwellian, but with temperature $e^{\beta t} T$ where $\beta = \alpha + 1/\tau_{esc}$. This clearly demonstrates that the effect of escape does {\it not} independently give a power-law distribution. 

In \citet{Guo2014,Guo2015}, it was concluded that an injection process is important for forming a power-law distribution. This can be shown by considering the injection term in Equation (3), illustrated in Fig.~\ref{fig:example}(a). We get
\begin{eqnarray} 
e^{\beta t} \left[\frac{d f}{d t} +  \beta f \right] = \frac{d}{dt} [e^{\beta t} f] = \frac{e^{\beta t}f_{inj}}{\tau_{inj}} 
\end{eqnarray}

Integrate Equation (5) along characteristics. If the injected energy spectrum is $f_{inj} = \frac{2 N_{inj}}{\sqrt{\pi}} \varepsilon^{1/2} \exp{(-\varepsilon)}$ and we inject particle continuously with this initial distribution from $t = 0$ to $t = \tau_{inj}$ with particle number $N_{inj} \propto V_{in}\tau_{inj}$ from the upstream ($\tau_{inj}$ is the time scale for particle injection and $V_{in}$ is the reconnection inflow speed), then










\begin{eqnarray} 
 &f& (\varepsilon, t) = 
 \frac{2 N_{inj} }{\sqrt{\pi} \alpha \tau_{inj} \varepsilon^{\beta / \alpha}}  [\Gamma_{1/2 + \beta/\alpha}(\varepsilon e^{-\alpha t}) - \Gamma_{1/2+\beta/\alpha}(\varepsilon))]
\end{eqnarray}

where $\Gamma_s(x)$ is the upper incomplete Gamma function. 
 Fig.~\ref{fig:example}(b) shows the solution as Equation (6) for different values of $\alpha \tau_{inj}$ (assuming no escape). As $\alpha \tau_{inj}$ increases, a power-law distribution forms and extends to larger and larger Lorenz factor.

We make several remarks on the issue of power-law formation. First, an escape effect is not necessary for forming a power-law energy distribution, or in other words, the escape term can be zero, and the Fermi acceleration gives a ``-1'' spectrum. The injection term is essential for developing a power law. However, note that it is still important to consider escape for determining the eventual shape of the spectrum in a realistic system, especially if both acceleration rate and escape rate have an energy dependence. This can be naively understood as that the spectral index $p$ is determined by the classical solution of Fermi acceleration $p = 1/(\alpha (\varepsilon) \tau_{esc}(\varepsilon)) +1$. The energy dependence in acceleration and escape needs to be combined to give a power-law distribution. Stochastic acceleration in turbulence has considerable difficulty to produce power-law distributions, as it needs to fine tune the energy dependence to produce a power-law distribution \citep[see discussion in e.g.,][]{Matthews2020}. \citet{Drury_2012} has illustrated that in a simple leaking box model, the escape term due to advection of energetic particles in the reconnection outflow can lead to escape that allows a power-law distribution. It is also worth noting that the form of the injected particle distribution does not decide the final distribution \citep{Guo2019}. At high energies, the acceleration process is determined by the Fermi process. \citet{Guo2019} has shown that including a pre-acceleration process at the X-point does not change the power-law index in any significant way. This further shows that Fermi acceleration is a general and robust mechanism for producing power-law energy spectra as suggested by numerous space and astrophysical energization processes.

In principle, parallel electric fields can give a power-law distribution as well, as long as it has a power-law energy dependence $\dot{\varepsilon} \propto \varepsilon^{\delta}$ and the power-law slope becomes $p = \delta$ when the escape effect is ignored. This result can be obtained by solving Eq. (2) similar to Eq. (6) \citep[see also][for a general discussion]{Drury1999}. For a relativistic particle moving close to the speed of light, the energy gain rate in the diffusion region $q \vec{v} \cdot \vec{E}$ is nearly independent of energy, whereas for a nonrelativistic particle $q \vec{v} \cdot \vec{E} \propto \varepsilon^{1/2}$. We see evidence of this process in the early stage of PIC simulations in Fig. \ref{fig:pic_run}. However, based on various studies this mechanism is localized around the X-line region and only accelerates particles to a limited energy. Whether it can accelerate a large number of particles to high energy in systems of realistic spatial extension is highly questionable. 

\subsection{Outlook}
\label{subsec:outlook}

Despite the remarkable progresses made in particle acceleration in relativistic magnetic reconnection there are several unsettled issues in reconnection acceleration models. The current understanding is likely temporary, and may evolve as we learn more. Here we discuss a number of unresolved issues.

\subsubsection{Injection process}
\label{subsubsec:injection}

What determines the separation between thermal and nonthermal particle distribution is an important one. There have been recent studies on this \citep{Ball_2019,Kilian2020,Sironi_2019,Che2020}.
While the acceleration of particles by parallel electric field is not effective in accelerating particles to high energy, it can be a useful mechanism for accelerating low-energy particles \citep{Ball_2019,Sironi_2019}. However the role of perpendicular electric field can be important as well, as trajectory analysis shows that some accelerated particles experience very little energization from the non-ideal electric field \citep{Guo2019}.
\citet{Kilian2020} have illustrated that processes involving the perpendicular electric field $E_\perp$ is important, and even more important than processes related to $E_\parallel$. We emphasize here again that because several competing mechanisms exists, it is important to study this statistically and consider both possibilities without bias. It should be noted that both of the processes are usually mixed together in low-energy acceleration, while at high energy the Fermi process is likely the dominant one.

The injection problem is more commonly studied in shock acceleration problem but not in the reconnection or turbulence scenario. This is because particles need to move significantly faster than the speed of the shock front in the shock acceleration problem. This barrier, however, is much easier to overcome in the case of reconnection, as the reconnection generated flow usually has a speed less than the upstream Alfven speed. In other words, the injection threshold can be fairly low. \citet{Kilian2020} have also discussed that the contribution from $E_\perp$ and $E_\parallel$ can depend on the system size and simulation duration, suggesting that it is difficult to determine the relative contribution of the two in a large scale system. Within the size range that is accessible to fully kinetic simulations $E_\perp$ seems to become more important with increasing system size.

\subsubsection{Guide field dependence}
\label{subsubsec:guide_field}

Most earlier simulations in the relativistic regime have been focusing on the case of weak guide field. In real astrophysical systems, it is of course naturally expected that the guide field is usually finite, even strong in certain circumstance. 
The presence of a guide field could have a significant consequence in reconnection physics, particle acceleration, and associated high-energy emission.  Knowledge from the nonrelativistic regime has shown that a guide field impacts the relative importance of acceleration mechanisms. In the guiding center description, the curvature drift acceleration decreases in efficiency as the guide field increase, as particles need to follow along a longer field line (curvature decrease) and the outflow speed decreases $\dot{\varepsilon} \propto \vec{\textbf{V}}_{flow} \cdot \vec{\kappa}$. Moreover, the acceleration by curvature drift motion is reduced by the gradient-B drift, which gives a net cooling effect during reconnection. The combined acceleration can be described by fluid compression and shear  \citep{Li_2018a}. In the low-$\beta$ reconnection with a weak guide field, the compression effect is important for supporting Fermi acceleration \citep{Li2018Role,Li2018Large,Du_2018}. When a finite guide field is included, the compression is reduced. The fluid shear effect and parallel electric field become more important \citep{Li2018Role,Li2018Large,Li2019Particle}. While it is still difficult to determine the consequence of this in real, large-scale system, it is expected that in a strong guide field limit ($B_g/B_0 \gtrsim  1$), the acceleration becomes much weaker \citep{Dahlin2016,Montag2017}. 

\subsubsection{3D physics}
\label{subsubsec:3d_physics}

Recent advances in computational plasma physics has allowed us to explore reconnection in 3D. It has been established that 3D reconnection can spontaneously generate turbulence \citep{Daughton2011,yhliu13a,Guo2015,Li2019Formation}.
Whether and how 3D physics changes the reconnection rate is subject of a major debate. However, 3D instabilities and turbulence will almost certainly modify the 2D picture of particle acceleration in magnetic reconnection, where turbulence plays an important role in particle transport and perhaps acceleration in the reconnection region.
Recent 3D studies have shown that 3D effects can be important for efficient acceleration in reconnection \citep{Dahlin2017,Li2019Formation}. In 2D magnetic field configuration particles are trapped in magnetic islands due to restricted particle motion across magnetic field lines \citep{Jokipii1993,Jones1998,Giacalone1994}, and high-energy particle acceleration can be prohibited. Chaotic field lines and turbulence due to 3D evolution of oblique tearing instability \citep[e.g.,][]{Daughton2011,yhliu13a} and drift kink instability \citep[e.g.,][]{Guo2015} make particles leaving the flux rope and can lead to efficient transport of particles in the reconnection region, which is found to be important for further acceleration in the reconnection region \citep{Dahlin2017,Li2019Formation}. While in current 3D simulations the particle energy spectra do not have a strong difference compare to the 2D simulations, 3D effects will be unavoidable and become important in large scales, presumably when those simulations can be done in a larger domain and longer duration.

\subsubsection{Toward a large scale theory}
\label{subsubsec:large_scale}

While it is important to gain insight using kinetic plasma simulations, the conclusions reached in those simulations are mostly proofs of basic ideas and cannot be directly compared with astrophysical observations. It has been shown in some work that PIC simulations with estimated or assumed magnetization $\sigma_0$ can provide particle acceleration and radiation features similar to observations \citep{Cerutti2013,Guo2016,Sironi2016,Petropoulou2016,Zhang2018}, but it is important to keep in mind that these observational features occur in much larger spatial and time scales than conventional plasma simulations can model.
Using typical numbers in \citet{Ji2011}, the ratio between the system size and the skin depth $L/d_e \sim 10^8$ for solar flares, $\sim 10^{13}$ for pulsar wind nebulae, and $\sim 10^{17}$ for extragalactic jets. Because of the huge scale separation between the system size and kinetic scales, it is impractical for conventional kinetic simulation methods to model the whole problem in any foreseeable future. The solution to this is a large-scale model that contains basic acceleration physics learned from kinetic simulations. 
Because the main acceleration mechanism in the reconnection layer is the Fermi acceleration process in the motional electric field, there have been attempts for modeling particle acceleration during magnetic reconnection in a macroscopic system by neglecting acceleration due to the non-ideal electric field \citep{Li2018Large,Beresnyak2016,Zank2014,LeRoux2015,Drake2019}.
The injection term may be parameterized and further Fermi acceleration can be studied using a Fokker-Planck description. \citet{Li2018Role} have provided analysis showing that the acceleration of particles in PIC simulations of magnetic reconnection can be described by fluid compression and shear, as in the energetic particle transport theory \citep{Parker1965,Zank2014LNP}. \citet{Li2018Large} has solved a classical Parker transport equation in the velocity and magnetic fields generated from a high-Lundquist-number magnetohydrodynamic simulation. It readily shows that the simulations give power-law distributions with spectral index weakly depending on the diffusion coefficient. \citet{Drake2019} and \citet{Arnold2019} presented a set of equations where the guiding-center particles feed back on the MHD equations so the total energy of the system for fluid and energetic particles is conserved. This however has not included the effect of particle scattering in turbulence, which is expected to be important in 3D turbulent reconnection.  In the relativistic case, a series of work has established the transport theory for the case with relativistic flow \citep{Webb1985,Webb1989,Webb2019}. In another approach, these can be studied in the description of generalized Fermi acceleration \citep{Lemoine2019}, where particle energization is studied by following the momentum of particles through a sequence of local frames where local electric field vanishes.

\section{Magnetic Reconnection Physics}

Back in 1957, \cite{sweet58a} and \cite{parker57a} derived the first reconnection model using the framework of resistive-MHD, which states that the normalized rate ($R$) is equal to the aspect ratio of the diffusion region ($\delta/L$). A normalized rate ($R\equiv V_{in}/V_A$) is the inflow speed normalized by the characteristic Alfv\'en speed since the inflow speed ($V_{in}$) measures how fast the plasma inflow brings  magnetic flux into the diffusion region for reconnection. Unfortunately, the long diffusion region length ($L$) in this model results in a rate that is too slow to explain the time-scale of solar flares; in order to explain the flare observation the normalized reconnection rate ($R$) should be of an order of $\mathcal{O}(0.1)$ \citep{parker63a}. Seven years later, Petschek proposed an idea of ``wave-propagation'' in the direction normal to the current sheet, and derived a steady-state solution where the outflow exhaust is bounded by a pair of standing slow shocks \citep{petschek64a}. These standing shocks help convert magnetic energy and divert the inflow to outflow; thus, reconnection outflow exhaust is opened up, and the diffusion region is localized in length; the reconnection rate can thus be fast. As we discussed in the Introduction section, although a conclusive observational evidence of fast reconnection remains challenging to obtain in astrophysical systems, observations of astrophysical flares \citep{Abdo2011,Tavani2011,Uzdensky2011,Arons2012,Kumar2015} suggest the importance of rapid magnetic energy release through the re-organization of magnetic topology, in favor of fast reconnection. In order to apply the idea of reconnection in astrophysical systems, both the Sweet-Parker's model and Petschek's model were extended to the relativistic regime where the outflow speed can approach the speed of light in strongly magnetized plasmas. The associated Lorentz contraction was proposed to enhance the reconnection rate \citep{Blackman1994,lyutikov03a,Liu2015}, but was later challenged by a pressure-balance argument \citep{lyubarsky05a}. 

While Petschek's open exhaust model remains a valid steady-state solution of fast reconnection, a Petschek-type analysis does not really address {\it why} and {\it how} the diffusion region is localized in the first place; because a Petschek solution collapses to the Sweet-Parker solution in uniform resistive-MHD simulations \citep{biskamp86a, sato79a}. This prompted researchers look into kinetic descriptions of dissipation that is beyond a simple, uniform resistivity \citep{Birn2001, rogers01a, shay99a, hesse11a, hesse99a, cassak05a,mandt94a,yhliu14a,stanier15a,tenbarge14a,daughton09a}, and search for the missing {\it localization mechanism}. In this section, we will review the feature newly observed in relativistic reconnection \citep{Liu2015}, which leads to a general reconnection rate model that provides the upper bound rate of an order $\mathcal{O}(0.1)$ in both the relativistic and non-relativistic limit \citep{Liu2017}. In the end, we review our new understanding on the localization mechanism in relativistic regime, which brings the system to the state of fast rate, and also explains the bursty nature during relativistic reconnection \citep{Liu2020}. As an example, we will show a PIC simulation of relativistic antiparallel reconnection in electron-positron plasma where the upstream magnetization parameter is $\sigma_{x} = B_x^2/(8 \pi w)=89$.

\subsection{Scale-separation in the ideal region} 
\begin{figure}
\centering
\includegraphics[width=0.65\textwidth]{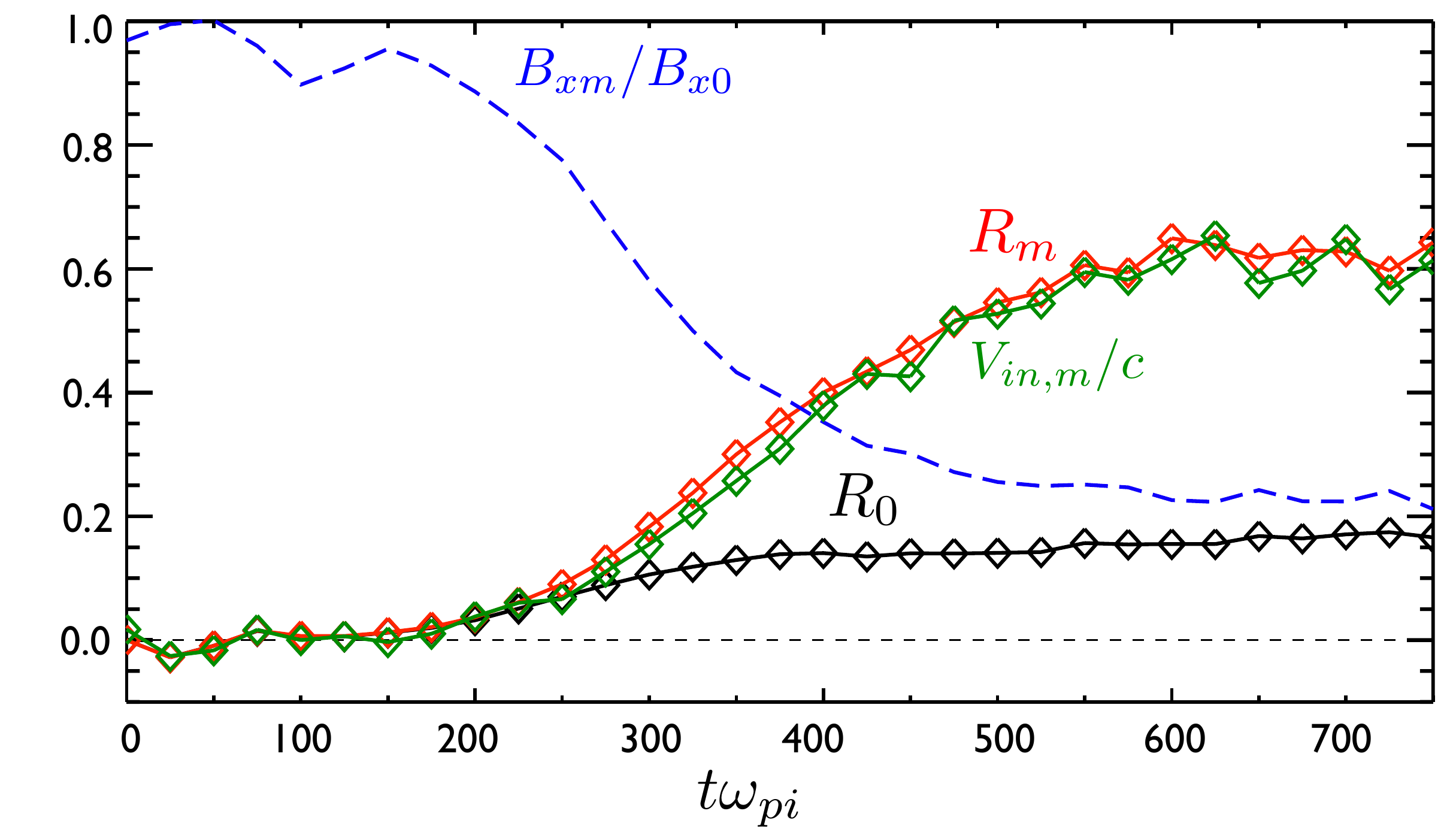}
\caption{Time evolution of several key quantities measured during a relativistic reconnection with upstream magnetization parameter $\sigma_{x} = 89$: $R_0$ is the reconnection rate normalized to the asymptotic values of quantities in macroscopic scale. $R_m$ is the rate normalized to quantities immediately upstream of the diffusion region (in microscopic scale). $V_{in,m}$ and  $B_{xm}$ are the inflow speed and the reconnecting magnetic field immediately upstream of the diffusion region (in microscopic scale), respectively \citep{Liu2017}. }
\label{fig:rate}
\end{figure}

In addition to the significant particle acceleration during relativistic reconnection, simulations in this extreme regime provide important insights into our understanding of reconnection physics, especially the long-standing reconnection rate problem. Fig.~\ref{fig:rate} shows the evolution of some key quantities relevant to the study of reconnection rate, including the inflow speed ($V_{in,m}$), reconnecting field strength ($B_{xm}$) immediately upstream of the diffusion region, reconnection rate normalized to the quantities immediately upstream of the diffusion region ($R_m\equiv V_{in,m}/V_{Am}= cE_y/B_{xm}V_{Am}$) and reconnection rate normalized to the asymptotic quantities ($R_0\equiv V_{in,0}/V_{A0}= cE_y/B_{x0}V_{A0}$). Here the subscript ``$m$'' means quantities at the microscopic scale immediately upstream of the diffusion region; ``0'' means the asymptotic value at the larger scale. Reconnection reaches the non-linear state after $\sim 400/\omega_{pi}$ in this case.

Interestingly, the inflowing plasma can reach a significant fraction of the speed of the light ($c$) before arriving at the diffusion region, as shown as $V_{in,m}$ in green. The ratio of this inflow speed to $c$ is basically the reconnection rate normalized to the quantities immediately upstream of the diffusion region, $R_m=V_{in,m}/V_{Am}\simeq V_{in,m}/c$, and it goes up closer to unity in the large $\sigma$ limit \citep{Liu2015}. Nevertheless, even though this ``micro-scale reconnection rate'' can go up to an order of $\mathcal{O}(1)$, the ``global reconnection rate'', $R_0\equiv cE_y/B_{x0}V_{A0}$, (as shown by the black curve in Fig.~\ref{fig:rate}) is still limited to the canonical value of an order of $\mathcal{O}(0.1)$ \citep{Liu2017, Sironi2016, zenitani08b}, as in the non-relativistic regime \citep{cassak17a, shay99a,hesse99a,daughton07a,bessho05a,swisdak08a}. This difference in $R_0$ and $R_m$ is due to a significant reduction of reconnecting field at the micro-scale, $B_{xm}$, as shown by the blue dashed curve in Fig.~\ref{fig:rate}. In other words, the reconnecting magnetic field at the microscopic scale can be different from the asymptotic field at the macroscopic scale. In light of the observation of this ``scale-separation'' at the region upstream of the diffusion region in the relativistic limit, we were able to develop a simple model to explain the fast global rate of value $\mathcal{O}(0.1)$ that is commonly observed in both the relativistic and non-relativistic regimes \citep{Liu2017}. \\

\subsection{A general model of magnetic reconnection rate} 
\begin{figure}
\centering
\includegraphics[width=0.86\textwidth]{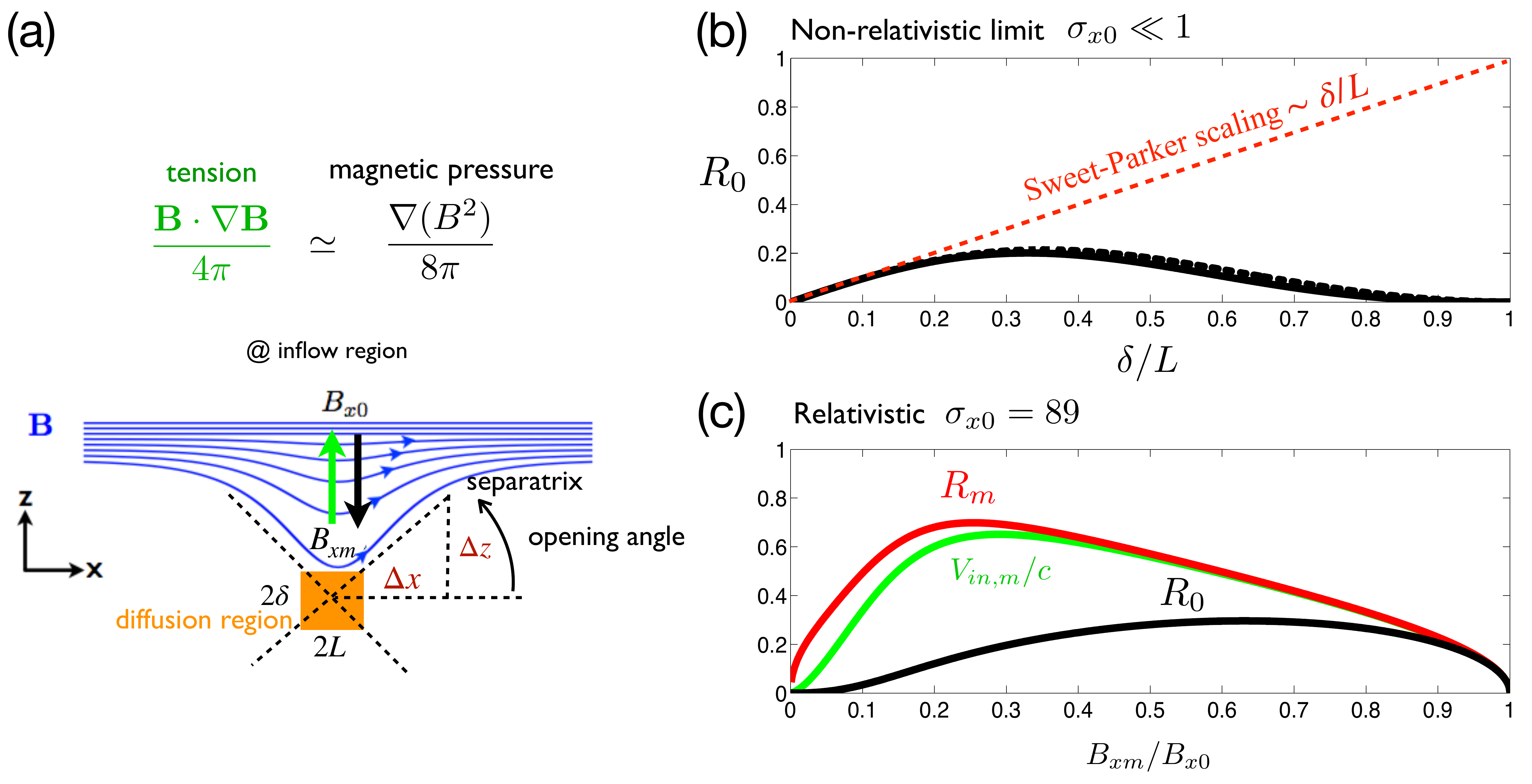}
\caption{The reconnection rate model: Panel (a) shows the upstream field line geometry, the dimension of the diffusion region and terms that are important in the force-balance. Panel (b) shows the predicted reconnection rate as a function of diffusion region aspect ratio $\delta/L$ in the non-relativistic limit. Panel (c) shows the predicted reconnection rate as a function of $B_{xm}/B_{x0}$ in the relativistic limit \citep{Liu2017}}.
\label{fig:model}
\end{figure}

The idea is surprisingly simple and general. Per Sweet-Parker scaling, the reconnection rate is basically the aspect ratio of the diffusion region (orange box in Fig.~\ref{fig:model}(a)) $\delta/L$, where $\delta$ and $L$ are the half-thickness and -length of the diffusion region, respectively. However, in the large $\delta/L$ limit (i.e., a localized diffusion region), the upstream magnetic field is indented that unavoidably induces a magnetic tension force pointing to the upstream region, as illustrated by the green arrow in Fig.~\ref{fig:model}(a). In the low-$\beta$ regime, the only term that can counterbalance this tension force is the magnetic pressure gradient (black arrow) pointing toward the x-line, which requires the reduction of the reconnecting field when it is convected into the diffusion region. Note that this field reduction is illustrated by the lower ``line-density'' of the in-plane field lines, and this effect is not considered in a Sweet-Parker type controlled-volume analysis. Intuitively, this reduction of magnetic field that actually reconnects will reduce the reconnection rate, and it indeed occur in the $B_{xm}$ measurement in Fig.~\ref{fig:rate}.  A similar argument applied to the downstream region leads to the reduction of the outflow speed, which also constrains the reconnection rate in the large $\delta/L$ limit. As a consequence, global reconnection rate $R_0$ as a function of the opening angle can be derived, as shown in Fig.~\ref{fig:model}(b) and it has a maximum rate around 0.2. In contrast, the Sweet-Parker scaling is shown as red dashed line, which does not have a bound in the large $\delta/L$ limit.

To obtain this quantitative analytical model, we analyze the balance between the tension force and the magnetic pressure gradient force at the upstream in Fig.~\ref{fig:model}(a) by simply discretizing the force-balance equation along the inflow direction (i.e., the z-direction), ${\bf B}\cdot B_z/4\pi\simeq\partial_z B^2/8\pi$, that is well-justified in the low-$\beta$ plasma. After straightforward simple algebra that also considers the geometry \citep{Liu2017}, an expression of the reconnecting magnetic field strength immediately upstream of the ion diffusion region ($B_{xm}$), as a function of the slope of the magnetic separatrix ($\Delta z/\Delta x$), can be derived: 

\begin{eqnarray} 
B_{xm}\simeq \left[\frac{1-(\Delta z/\Delta x)^2}{1+(\Delta z/\Delta x)^2}\right] B_{x0} \quad.
\end{eqnarray}

This slope ($\Delta z/\Delta x$) quantifies the opening angle ($\simeq\mbox{tan}^{-1}(\Delta z/\Delta x)$) of reconnection exhausts that gives rise to the upstream tension force, and it is essentially the diffusion region aspect ratio $\delta/L$. The predicted reconnecting magnetic field monotonically decreases to 0 when the slope ($\Delta z/\Delta x$) approaches unity. A similar analysis of the force balance including the plasma inertia, $nm_i {\bf V}\cdot \nabla {\bf V}$, inside the diffusion region along the outflow direction, gives the outflow speed:

\begin{eqnarray} 
V_{out,m}\simeq c \sqrt{\frac{(1-\delta^2/L^2)\sigma_{xm}}{1+(1-\delta^2/L^2)\sigma_{xm}}} \quad.
\end{eqnarray}

In this calculation the relativistic effect on the Alfv\'en speed is included as the $\sigma_{xm}$-factor. The speed that transports reconnected flux out of the diffusion region is reduced in the $\delta/L$ $(\simeq \Delta z/\Delta x)\rightarrow 1$ limit. Combining these two quantities, we obtain the reconnection electric field $E_y\simeq B_{zm}V_{out,m}/c$, where $B_{zm}\simeq B_{xm} \delta/L$ is derived using $\nabla\cdot {\bf B}=0$; A large exhaust opening angle ($\simeq$ diffusion region aspect ratio) therefore leads to the reduction of reconnection rate $R_0=cE_y/B_{x0}V_{A0}$. The analytical predictions of $R_0$, $R_m$ and $V_{in,m}/c$ are then plotted as a function of $B_{xm}/B_{x0}$ for a relativistic run with $\sigma\simeq 89.0$ in Fig.~\ref{fig:model}(c). For $B_{xm}/B_{x0}\simeq 0.22$ as measured in Fig.~\ref{fig:rate}, the predicted $R_0=0.14$, $R_m=0.69$ and $V_{in,m}=0.62c$, and they explain well the plateau values in Fig.~\ref{fig:rate}, that corresponds to the quasi-steady state.

The upper bound value of reconnection rate $R_0\equiv cE_y/B_{x0}V_{A0}\sim$ 0.2 is then derived by combining these two effects, and a wide range of opening angles indicate a similar rate of order $\mathcal{O}(0.1)$, as shown in Fig.~\ref{fig:model}(b). This explains why the reconnection rate $\sim \mathcal{O}(0.1)$ is commonly observed in simulations and is often directly measured or indirectly inferred in Earth's magnetosphere, solar flares, and laboratory plasmas [e.g., reviewed in \cite{cassak17a}]. This nearly universal rate is essentially an upper bound value provided by the constraints (i.e., force-balance) imposed at the inflow and outflow directions. This analytical approach also works for asymmetric reconection \citep{yhliu18a}, the derived maximum rate is consistent with the analytical model \citep{cassak07b, cassak08a} often used to compare with magnetopause reconnection. Recent in-situ observations of NASA's Magnetospheric Multiscale (MMS) mission \citep{burch16a} has explored this nearly universal fast rate of collisionless magnetic reconnection in Earth's magnetotail \citep{TKMNakamura18a,genestreti18a,tobert18a} and magnetopause \citep{LJChen17a,burch16b}, and found a consistent fast rate around $\mathcal{O}(0.1)$. Furthermore, the high time-cadence and tight tetrahedron formation of MMS's four identical satellites enables scientists study the breaking mechanism of the frozen-in condition \citep{egedal19a} and the nature of reconnection electric field \citep{RNakamura18a}, that also appear to be consistent with the theory (e.g., \cite{hesse99a,hesse11a} and references therein). 
\\ 

\subsection{The need of a localization mechanism} 
The question is then what brings the system into the state of fast reconnection, where the diffusion region is localized, not extending to the system size; i.e., the large $\Delta z/\Delta x$, or small $B_{xm}/B_{x0}$ solution in Fig.~\ref{fig:model}(b). The answer may differ in different systems \citep{yhliu18c}. Here we identify the {\it localization mechanism} in the strongly magnetized regime, which also explains the bursty nature of relativistic reconnection.

\begin{figure}
\centering
\includegraphics[width=0.85\textwidth]{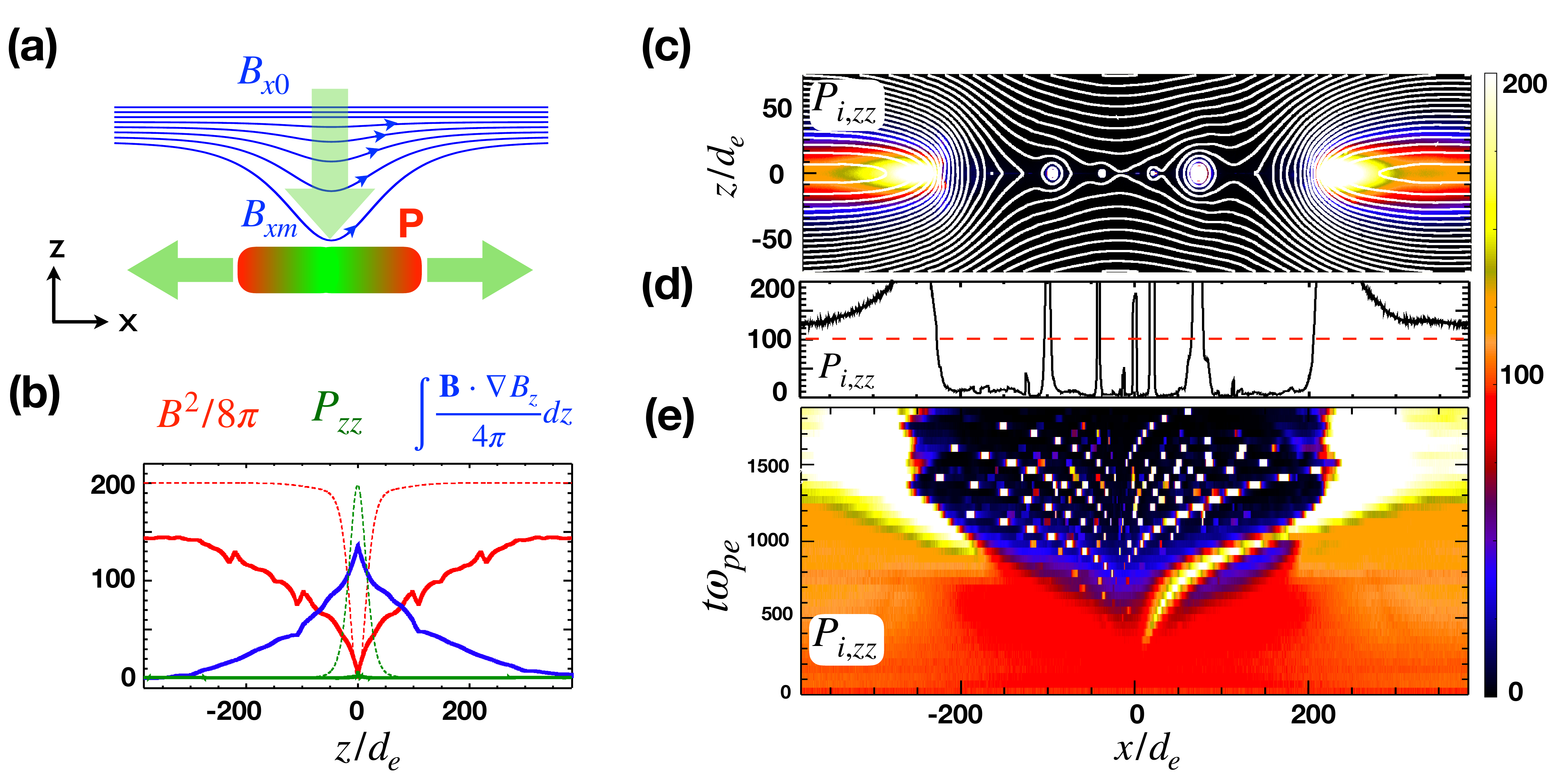}
\caption{Pressure depletion as the localization mechanism: Panel (a) illustrates that inflowing low-pressure plasma act to deplete the pressure right at the x-line, which can lead to the opening of exhausts. Panel (b) shows the critical role of tension force in balancing the force at late time (solid lines). The initial profiles are plotted as dashed lines for comparison. Panel (c) shows the positron pressure $P_{i,zz}$ that is relevant to the force-balance in the z-direction. Panel (d) shows the cut of $P_{i,zz}$ along the symmetry line at $z=0$. The horizontal red dashed line marks the initial value . Panel (e) shows the time-stack plot of $P_{i,zz}$ cuts along the symmetry line at $z=0$. This panel shows the bursty nature of reconnecting current sheet. \citep{Liu2020}}
\label{fig:depletion}
\end{figure}

By furthering the approach laid out in the previous discussion, we gain an interesting new insight. As illustrated in Fig.~\ref{fig:depletion} (a), during relativistic reconnection, inflowing low-pressure plasma (green part) acts to deplete the high plasma pressure (red part) in the initial planar current sheet. If this pressure depletion cannot be overcome by the kinetic thermal heating inside the diffusion region, an elongated diffusion region is not a plausible steady-state solution. The {\it only} way to restore the force-balance along the inflow direction is to develop a localized geometry, as shown in Fig.~\ref{fig:depletion}(a); because the indented upstream magnetic field will then induce a tension force pointing upstream, balancing the upstream magnetic pressure. This pressure depletion is a localization mechanism needed for achieving fast reconnection \citep{Liu2020}. As long as some degree of localization presents in the system, the reconnection rate $R_0$ easily reaches the value around $\mathcal{O}(0.1)$ in Fig.~\ref{fig:model} (c). 

This pressure depletion is evident in strongly magnetized plasmas. As shown in Fig.~\ref{fig:depletion}(c), the pressure $P_{i,zz}$ at the x-line drops significantly by a factor of $\sim 100 \times$ from its initial value to a value closer to the upstream value in the nonlinear state. This is clearly seen in the cut of $P_{i,zz}$ along the symmetry line $z=0$ in Fig.~\ref{fig:depletion}(d) where the horizontal red dashed line marks the initial relativistic thermal pressure that is required to balance upstream magnetic pressure $B_{x0}^2/8\pi$ in a planar (i.e., non-localized) sheet. The resulting change of the force-balance across the x-line is shown in Fig.~\ref{fig:depletion}(b). Initially, the upstream magnetic pressure $B_{x0}^2/8\pi$ (red dashed) is balanced by the thermal pressure $P_{zz}$ (green dashed) of the relativistically hot plasmas inside the current sheet (as the initial setup of a stable planar sheet; i.e., note that this pressure depletion process also works in a current sheet that is initially force-free). The system evolves to a quasi-steady state, where $B_{x0}^2/8\pi$ (red solid) is balanced by the magnetic tension force $\int{({\bf B}\cdot\nabla B_z/4\pi) \mathrm{d}z}$ (blue solid) instead, since $P_{zz}$ (green solid) is basically depleted. This evolution is consistent with the picture in Fig.~\ref{fig:depletion}(a). The microscopic reconnecting field $B_{xm} \simeq \sqrt{8\pi P_{zz}(z=0)}$ is thus brought down significantly as seen in Fig.~\ref{fig:rate}, which leads to $R_m \rightarrow \mathcal{O}(1)$ as also expected in Fig.~\ref{fig:model}(c). The pressure component in the z-direction can not be sustained because the majority of the available magnetic energy is converted to the kinetic energy of current carriers in relativistic plasmas. Note that this pressure depletion is relevant to the pressure-balance argument in \cite{lyubarsky05a}. However, while \cite{lyubarsky05a} used the pressure-balance to argue against a faster rate (i.e., $R_0 \rightarrow \mathcal{O}(1)$) enhanced by Lorentz contraction, here we used it to reveal the crucial localization mechanism needed for fast reconnection. 

To have a stable open geometry, the plasma needs to be heated up to the initial value while it is convected out to the downstream (the red part in Fig.~\ref{fig:depletion}(a)), so that the force-balance across the exhaust is maintained; i.e., due to the geometry it can not be the tension force as at the x-line. However, the kinetic heating inside the exhaust of relativistic reconnection is still not efficient enough in heating up outflow plasmas, thus the low-pressure region extends into the outflow, causing the once opened exhaust to collapse. The collapsing exhaust triggers the growth of secondary tearing islands, which helps balance pressure but only temporarily before they are ejected out (by outflows from the primary x-line at the center). The system can not reach a true equilibrium but rather a dynamical balance characterized by repetitive generation of secondary islands, as clearly seen in the time-stack plot of $P_{i,zz}$ cuts in Fig.~\ref{fig:depletion}(e). This pressure collapse also repeatedly excites concentric shock waves, whose signature can be seen as spikes in the $B^2/8\pi$ (red) curve of Fig.~\ref{fig:depletion}(b). Getting to the bottom of the bursty nature of relativistic reconnection, we conclude that this pressure depletion is responsible for triggering copious secondary islands, likely not the plasmoid-instability \citep{loureiro07a} derived from a force-balanced current sheet in resistive-MHD model. \\

\subsection{Some comments on the effect of guide field} 
While anti-parallel reconnection may trigger the most energetic events observable in nature, reconnection with a finite guide field (out of the reconnection plane) can be more common. The total energy release from an ensemble of guide-field reconnection events could be significant compared to a singular anti-parallel event. With a finite guide field, the in-plane Alfv\'enic speed $V_{Ax}=c\sqrt{\sigma_x/(1+\sigma_x+\sigma_g)}$ is the projection of the total Alfv\'en speed  $V_A=c\sqrt{\sigma/(1+\sigma)}$ to the reconnection plane \citep{Liu2015, melzani14a}, where $\sigma_g =B_g^2/(8\pi w)$ is the contribution from the guide field component. Thus, no matter how large the total $\sigma=\sigma_x+\sigma_g$ is, the reconnection outflow speed becomes (at most) mildly relativistic only (i.e., $\Gamma_{out} \simeq \sqrt{(\sigma+1)/(\sigma_g+1)} \simeq 1$) as long as $\sigma_g \gtrsim \sigma_x$. However, the normalized rate $R_0$ is expected to remain an order of $\simeq \mathcal{O}(0.1)$ if it is properly normalized to the in-plane Alfv\'enic speed, as seen in \cite{Liu2015} (where $R_0\simeq R_m$ in this limit). In addition, recent simulations of guide field reconnection in the relativistic regime \citep{Ball_2019,Rowan_2019,Liu2020} reveal a single x-line geometry that is more stable compared to the anti-parallel case (which appears to be bursty with repetitive generation of plasmoids as in Fig.~\ref{fig:depletion}(e)). This morphology difference can be explained by the rotation of the reconnected field out of the reconnection plane in the guide field case \citep{levy64a,lin93a,lyubarsky05a,yhliu11b}, since the enhanced magnetic pressure from the out-of-plane component $B_y^2/8\pi$ help balance the pressure across exhausts, preventing the open exhaust from collapsing back \citep{Liu2020}.







\section{Summary}

Recently magnetic reconnection in the relativistic regime has been studied actively, providing a viable explanation to fast energy release and high-energy emission in high-energy astrophysics. 
We summarize some recent progress on relativistic magnetic reconnection, with more discussion on several important theoretical issues such as particle acceleration mechanism, power-law formation, as well as reconnection physics including the rate problem. These advances have provided important basis for applying relativistic magnetic reconnection in high-energy astrophysical phenomena.

We reviewed the major particle acceleration mechanisms and how power-law energy distributions are formed in magnetic reconnection. Reconnection in magnetically dominated, relativistic plasmas provides an unique opportunity for studying the formation of power-law distributions because of the strong energy conversion and the associated particle acceleration. Recent PIC simulations have shown that power-law particle energy spectra emerges in relativistic magnetic reconnection and the spectral index approaches $p=1$ when the magnetization is large enough. By means of statistical analysis in PIC simulations, recent studies have agreed that Fermi mechanism rather than the parallel electric field is the dominant process for high-energy particle acceleration. We showed analytically that particle injection from the reconnection inflow is necessary for the formation of power-law. Particle escape is not necessary for the formation of power-law as previously claimed but it affects the resulting power-law index. The power-law index is found to be determined by the Fermi-like processes rather particle acceleration in the diffusion region.

A long-standing problem in reconnection studies is how magnetic energy is released and the time scale of this process. PIC simulations have shown that the normalized reconnection rate in relativistic reconnection is of $\mathcal{O}(0.1)$, which is similar to that of their non-relativistic counterparts. A general model has been constructed to quantitatively explain the rate within 2D current sheet geometry. Recent progresses feature the pressure depletion as the localization mechanism in the strongly magnetized regime. The pressure depletion leads to the collapse of the current sheet and the formation of multiple plasmoids in anti-parallel reconnection. In the relativistic regime, a finite guide field can prevent the collapse of the reconnection exhaust and thus results in fewer plasmoids, as recently found in PIC simulations of relativistic guide-field reconnection.

We pointed out several problems to be further studied for understanding the energy conversion and particle acceleration in reconnection. First, how particles are first injected into high energies from the thermal pool for further acceleration by the Fermi processes? We know that both parallel and perpendicular electric fields can inject particle, but it seems that their relative contributions depend on the guide field and the simulation box sizes. It is unclear what the major injection mechanism is in a large-scale system. Second, how does the guide field change the acceleration processes? Current understanding is that a finite guide field will modify the main acceleration mechanisms and reduces the efficiency of the Fermi-like processes, but these results are mostly based on PIC simulations of non-relativistic reconnection. A detailed study on this in relativistic reconnection is necessary. Third, what are the roles of 3D physics? 3D PIC simulations of reconnection have shown the generation of plasma turbulence, which enhances the transport of energetic particles. It is found that this can enhance particle acceleration in non-relativistic reconnection but seems to have limited effects on particle acceleration in relativistic reconnection. However, these statement are all based on a limited number of 3D simulations. To fully understand the roles of 3D physics, more 3D simulations are required. Last but not least, how to apply the physics learned from PIC simulations to explain large-scale astrophysical phenomenon? We pointed out that some kinds of large-scale models (e.g. energetic particle transport equation, guiding-center description) in non-relativistic/relativistic flows are necessary to account for the acceleration and radiation occurring in astrophysical scales. This requires the developments in both analytical models and numerical modeling.

We also note that the recent advances are very beneficial to the basic theories of particle acceleration and magnetic reconnection, in that it drives theoretical understanding in these extreme conditions. For example, the recent progress on particle acceleration in relativistic reconnection is very important for testing reconnection acceleration theory. In addition, earlier reconnection kinetic physics, especially reconnection rate, has been tested and developed in the extreme relativistic condition.

\section*{Acknowledgements}
We gratefully acknowledge discussions 
with Joel Dahlin, Jim Drake, Quanming Lu, Lorenzo Sironi,  Anatoly Spitkovsky, and Marc Swisdak. 
We acknowledge support from DOE through the LDRD program at LANL and DoE/OFES support to LANL, and NASA Astrophysics Theory Program. F. G. and W. D. acknowledge support in part from NASA Grant 80NSSC20K0627. Y. L. and X. L. are supported by NSF-DoE 1902867 and NASA MMS 80NSSC18K0289. The research by P. K. was also supported by the LANL through its Center for Space and Earth Science (CSES). CSES is funded by LANL's Laboratory Directed Research and Development (LDRD) program under project number 20180475DR. Simulations were performed at National Energy Research Scientific Computing Center (NERSC) and with LANL institutional computing.

\section*{DATA AVAILABILITY}
Data sharing is not applicable to this article as no new data were created or analyzed in this study.

\bibliographystyle{aasjournal}

\end{document}